# CORRELATION BETWEEN INTERNAL STATES AND STRENGTH IN BULK METALLIC GLASS

J. Tan[1, 2, 3,*], C.J. Li[1, 2], Y.H. Jiang[1], R. Zhou[1], J. Eckert[2, 3],

[1]School of Materials Science and Engineering, Kunming University of Science and Technology, 650093 Kunming, China
[2]IFW Dresden, Institute for Complex Materials, P.O. Box 27 01 16, D-01171 Dresden, Germany
[3]TU Dresden, Institute of Materials Science, D-01062 Dresden, Germany

Keywords: bulk metallic glass, free volume, strength, internal states, glass transition

## Abstract

The internal states or local structures of bulk metallic glass (BMGs) can be well reflected from the changes of density, structural relaxation as well as the elastic constants. With the increasing free volume (FV) content, more local atomic clusters are capable of simultaneous plastic shear at different sites in the metallic glasses, inducing large plasticity. In this work, we report a close correlation between the internal states and strength in a BMG and discover that the yield strength can be changed by varying of the casting current, revealing that the yielding strength of BMGs is not only intrinsically associated with the glass transition, but also with the internal states, such as free volume and elastic properties. Such results may have some implications for understanding the correlations between the internal states and mechanical properties of BMGs.

## 1. Introduction

The rapid solidification for quenching metallic liquids at very high rates allows freezing of the configuration of alloy melts and obtaining bulk metallic glass (BMG) for some alloys, depending on the glass-forming ability (GFA) of the respective alloy [1]. The rapid solidification processing results in increased strength, high elastic strain limit (up to 2%), and relatively low Young's modulus as well as enhanced fracture and fatigue resistance, with concurrent improvements in mechanical, physical and chemical properties [2, 3]. The deformation of BMG at room temperature is highly localized into shear bands, owing to strain softening and/or thermal softening [4, 5]. Finally, they develop towards catastrophic failure along one dominant shear plane, thus restricting wide application of BMG [6]. Therefore, further understanding the relationship between mechanical properties and internal states might assistant the design and development of BMGs and their composites with practical engineering application [7].

The internal states or local structures of bulk metallic glass (BMGs) can be well reflected from the changes of density, structural relaxation as well as the elastic constants [8-10]. The shear modulus ($\mu$) and bulk modulus ($B$) are representative of the change in size and shape of materials, respectively. Hence, the Poisson's ratio ($v$) or the ratio of bulk-to-shear modulus ($B/\mu$) correlates with the toughness of BMGs and materials with different Poisson's ratios behave mechanically very different. The properties range from 'rubbery' to 'dilatational', between which are 'stiff' materials like metals and minerals, 'compliant' materials like polymers and 'spongy' materials like foams[11].

---

[*] Corresponding author at: School of Materials Science and Engineering, Kunming University of Science and Technology, 650093 Kunming, China. Tel: +86 871 65107512; Fax: +86 871 65107922; Email address: tanjuncn@gmail.com (J. Tan)

As a matter of fact, the combination of both strength and plasticity is a vital requirement for most structural materials [12-18]. In our previous research [19-21], we found a close correlation between the internal states and plasticity in a BMG by tuning the casting current. Optimization of the current can induce large plasticity in an otherwise brittle BMG. The larger plasticity corresponds to the internal states with more average FV as revealed by lower density, higher enthalpy change and higher Poisson's ratio $v$ (or the modulus ratio $B/\mu$). To further investigate the relationship between the internal states and mechanical properties, furthermore, the direct correlation between internal states and strength of the BMGs also arouses our concerns [22-27].

## 2. Experimental procedures

### 2.1 Preparation of the alloys

Fully amorphous $Zr_{56}Co_{28}Al_{16}$ BMG samples were obtained employing different electrical power of the arc by controlling the discharge current during casting (labeled as samples S1, S2, S3, and S4 obtained with gradually increasing current, respectively). $Zr_{56}Co_{28}Al_{16}$ (at. %) pre-alloys were synthesized by arc-melting the proper amounts of the constituting elements (purity of 99.99% or higher) in a Ti-gettered high purity argon atmosphere. The pre-alloy ingots were re-melted at least four times in order to ensure chemical homogeneity. The ingots were then cut into small pieces with a mass of around 5-6 g, and then they were melted and cast into cylindrical rods with a diameter of 2 mm and a length of approximate 50 mm using water cooling copper mold suction casting technique.

### 2.2 Characterizations of the BMGs

The structure was checked by x-ray diffraction (PANalytical X'Pert PRO Diffractometer, Co radiation, reflection geometry). The micro-indentation was carried out at least 15 points on each of the samples using a Vickers diamond indenter and a 25 g load with a dwell time of 10 s. Compression tests were performed on a universal tension/compression testing machine (Instron 5869) equipped with a laser extensometer (Fiedler) at a constant strain rate of $2.5 \times 10^{-4} s^{-1}$ at ambient temperature. The samples used in the compression tests were prepared as follows: firstly, small cylinders with a height of approximately 4.2 mm were cut from the as-cast 2 mm diameter rod using a wire saw, and then they were carefully polished to achieve a height to diameter ratio of 2:1 and a good parallelism between the top and bottom end. Compression tests were performed at least five to six times using the samples cut from similar positions of different as-cast rods, and the results turned out to be rather reproducible. The elastic properties were determined with an Olympus Panametrics-NDT 5900PR ultrasonic testing device. Here, utilizing a Metter Toledo UMT2 electrical balance with a sensitivity of 0.1 μg, the densities of the samples were measured with the help of in-house designed apparatus using the Archimedean principle. The effective resolution was 0.0005 g/cm$^3$.

## 3. Results and discussions

### 3.1 Thermodynamic origin of shear band formation and yielding

A universal scaling law from thermodynamics in BMGs has been derived and uncovers the thermodynamic origin of shear band formation and yielding by [28-30]:

$$\tau_y = 3R(T_g-T_r)\rho_r/M, \qquad (1)$$

Where $R$ is the gas constant, $T_g$ is the glass transition temperature, $T_r$ is room temperature, $\rho_r$ is the density at room temperature and $M$ is the molar mass of the material. From this, it can be seen that the yielding of BMGs by the formation of multiple shear bands is closely related to the glass transition, the density and the molar mass of materials.

## 3.2 Quantification of the free volume

Free volume (FV) can be created by the inelastic deformation of local atomic clusters under shear stress or annihilated by atomic diffusion and/or atomic jumps during the process of casting [31-33]. Thus it can be viewed as an effective method to measure the internal states in MGs. In the FV model [34-40], the absolute value of the FV contained in MGs at a given temperature is, however, difficult to measure because it is almost impossible to obtain a reference amorphous state without any FV in a specific experiment. Many attempts, such as, the specific heat capacity [41], the density change [19, 42, 43], the positron annihilation lifetime [44, 45] and the thermal dilatation test methods [46], have been utilized to estimate the content of FV. Among them, the density change method is the most convenient and effective method through structural rearrangement to evaluate the FV difference in the glassy structure [42, 43]. The reduced FV ($x$) can be quantitatively expressed by [19, 42, 43, 47] :

$$x \approx 2(\rho_o/\rho_r - 1), \qquad (2)$$

$\rho_0$ is the density of ideal glass, which is usually approximated as the sufficiently crystallized samples [19, 43]. Thus, $\rho_r$ can be written as:

$$\rho_r \approx 2\rho_o/(2+x). \qquad (3)$$

## 3.3 The influence of free volume on the yield strength

Plastic deformation in metallic glasses is possibly interpreted as the thermally activated creation and rearrangement of shear transformation zones (STZs) based on the FV model [31-33]. It is noted that, the equations above illustrate that the density of metallic glass is less than that of ideal glass. So, $\tau_y$ can be further correlated with the reduced FV:

$$\tau_y \approx 6R\rho_o(T_g-T_r)/[M(2+x)]. \qquad (4)$$

## 3.4 Verification of the model

To verify the model above, fully amorphous $Zr_{56}Co_{28}Al_{16}$ BMG samples were obtained employing different electrical power of the arc by controlling the discharge current during casting (labeled as samples S1, S2, S3, and S4 obtained with gradually increasing current, respectively) [19]. Fig. 1 shows X-ray diffraction patterns of the as-cast samples in a rod form of 2 mm in diameter. The patterns of all the samples, taken from the cross-sectional surfaces of the as-cast rods, exhibit typical broad diffraction maxima without any detectable crystalline Bragg peaks, indicating that all samples are fully amorphous and there are no obvious differences within the resolution limits of XRD.

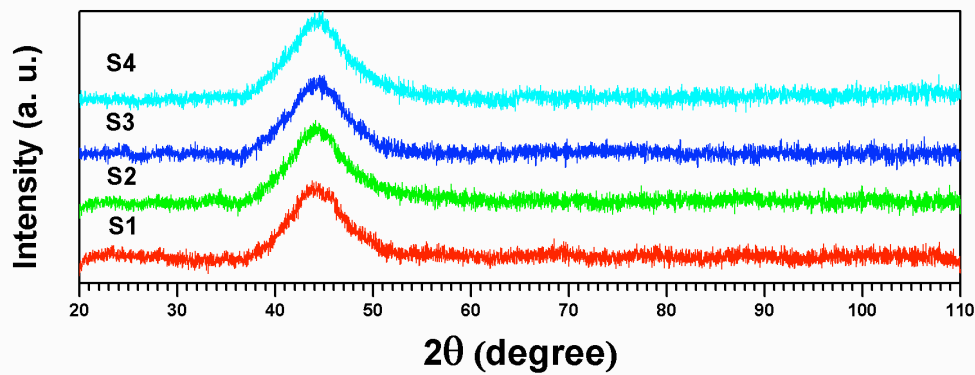

Fig. 1 XRD patterns of as-cast $Zr_{56}Co_{28}Al_{16}$ BMG samples cast under different current.

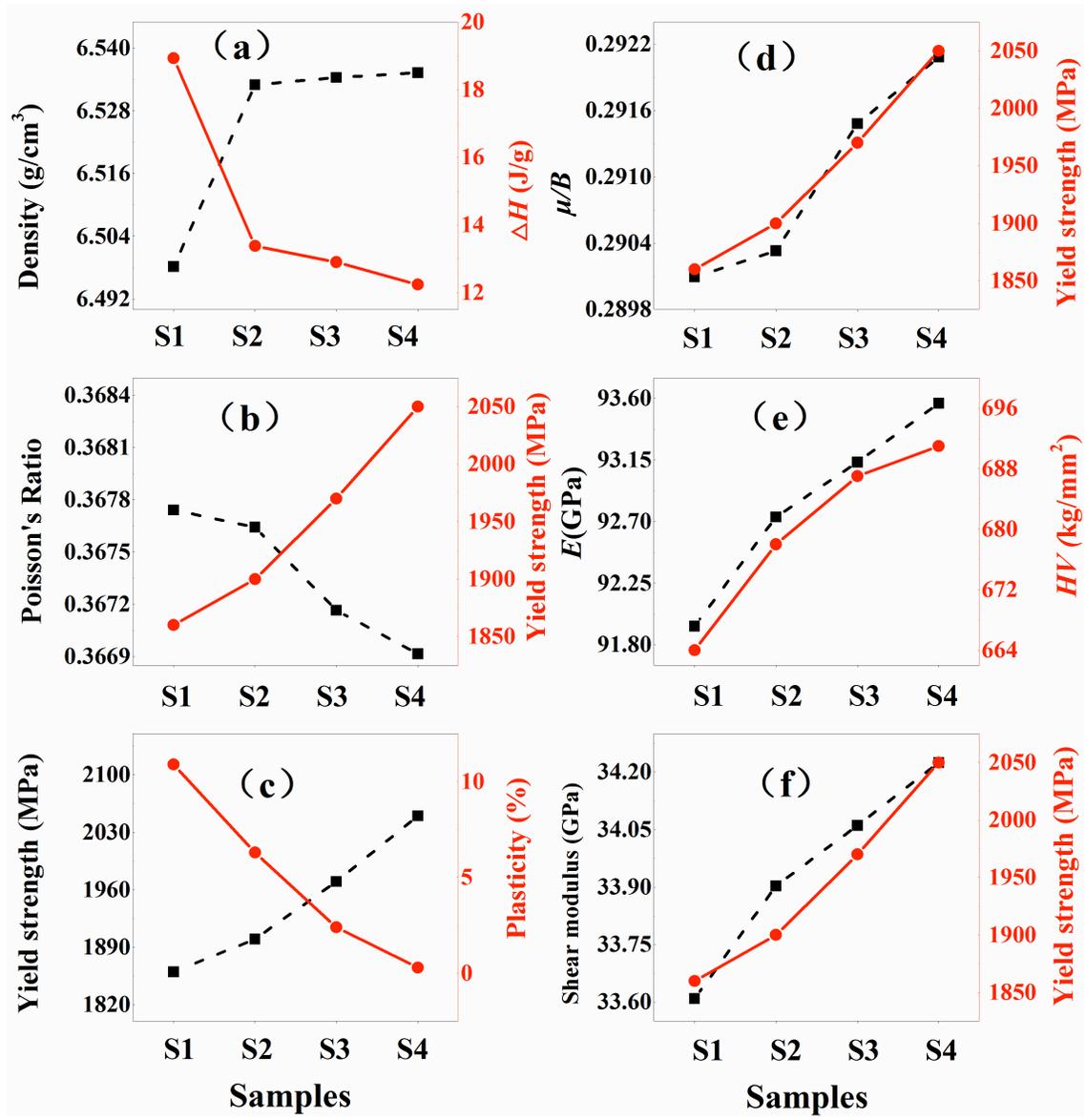

Fig. 2 Correlations between internal states and mechanical properties of as-cast $Zr_{56}Co_{28}Al_{16}$ BMG samples cast under different current. (a) Density and enthalpy change ($\Delta H$), (b) Poisson's ratio ($v$) and yield strength, (c) Yield strength and plasticity, (d) the ratio of shear-to-bulk modulus ($\mu/B$) and yield strength, (e) Young's modulus ($E$) and Vikers hardness ($HV$), (f) Shear modulus and yield strength as a function of different samples .

Fig. 2 depicts the correlations between internal states and mechanical properties of as-cast $Zr_{56}Co_{28}Al_{16}$ BMG samples cast under different current. With the increasing casting current, the density of the samples increases; while the enthalpy change ($\Delta H$) decreases, as shown in Fig. 2a. It is well known that the exothermic event prior to $T_g$ strongly links with the existence of free volume ($\Delta v_f$) in BMGs according to $\Delta H = \beta \, \Delta v_f$ [48], where $\beta$ is a constant and $\Delta v_f$ is the change of free volume per atomic volume. So the distinguishable drops of the $\Delta H$ indicate the decrease of the free volume of the as-cast samples with the power increasing. With the decreasing Poisson's ratio ($v$), the yield strength decreases, shown in Fig. 2b. The attainment of both strength and toughness is a vital requirement for most structural materials; unfortunately these properties are generally mutually exclusive [49]. When the yield strength increases, the plasticity dramatically drops to almost 0 (shown in Fig. 2c). With the increasing strength, the ratio of shear-to-bulk modulus ($\mu/B$), Young's modulus, Vikers hardness ($HV$) and shear modulus increases too, illustrated in Fig. 2d, 2e and 2f.

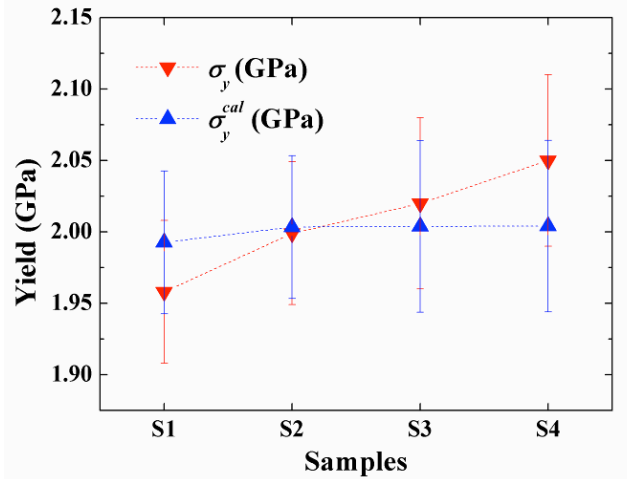

Fig. 3 Yield strength as a function of different samples casted under different casting current for $Zr_{56}Co_{28}Al_{16}$ BMG alloy indicating remarkable agreement between the calculated ($\sigma_y^{cal}$) and measured ($\sigma_y$) strength values.

Therefore, one can see that the sample with a minimum value of the density shows large excess FV ($x = 2.31\%$). With increasing density, the FV dramatically decreases from 1.23% to 1.16%. The sample S1 with minimum value of the density exhibits the largest critical fracture strain of about 12.8 %, and the samples S2 and S3 show lower plastic deformation ability of 7.8 % and 4.0 %, respectively. The sample S4, however, presents almost no plastic deformation before failure [19]. The calculated yield stresses agree well with the data obtained from the experiments, as shown in Fig. 3. This indicates that optimization of the arc-melting current during suction casting can impact the yielding strength of BMG. The higher strength corresponds to the internal states with less average FV, as revealed by higher density. From Eq. (3), it can be deduced that $\tau_y$ decreases when $x$ increases. On the other hand, with increasing FV content, the plasticity improves dramatically. Thus, more local atomic clusters, which are typically conceived to be about 5 atomic diameters in size [31], are capable of simultaneous plastic shear at different sites in the metallic glasses with slightly increased FV content and multiple shear bands are formed subsequently, inducing large plasticity [31-33]. Therefore, a higher FV content makes it easier to form multiple shear bands while, unfortunately, leading to lower yield stress.

## 4. Conclusions

With the increasing free volume (FV) content, more local atomic clusters are capable of simultaneous plastic shear at different sites in the metallic glasses, inducing larger plasticity but lower yield strength. The close correlations between the internal states and strength in a BMG have been summarized and it is discovered that the yield strength can be changed by varying of the casting current, revealing that the yielding strength of BMGs is not only intrinsically associated with the glass transition, but also with the internal states, such as free volume, density, enthalpy change ($\Delta H$), Poisson's ratio ($v$), the ratio of shear-to-bulk modulus ($\mu/B$), Young's modulus, Vikers hardness ($HV$) and shear modulus $etc$. Such results may have some implications for understanding the correlations between the internal states and mechanical properties.

## 5. Acknowledgements

The authors thank S. Donath, M. Frey, B. Opitz, U. Wilke, B. Bartusch, Y. Zhang, M. Stoica and U. Kühn for technical assistance and stimulating discussions. C. J. L and J. T would like to acknowledge the fellowship support from the China Scholarship Council. Additional supports provided by the German Science Foundation under the Leibniz Program (Grant EC 111/26-1), the Science Foundation of the Yunnan Provincial Science and Technology Department (Grant 2011FB021) and the Science Foundation of the Yunnan Provincial Education Department (Grant 2010Z012) are also gratefully acknowledged.